\documentclass{CUP-JNL-DAP}%

\usepackage{graphicx}
\usepackage{multicol,multirow}
\usepackage{amsmath,amssymb,amsfonts}
\usepackage{mathrsfs}
\usepackage{amsthm}
\usepackage{apacite}
\usepackage{rotating}
\usepackage{appendix}
\usepackage[authoryear]{natbib}
\usepackage{ifpdf}
\usepackage[T1]{fontenc}%
\usepackage{times}%
\usepackage{newtxmath}%
\usepackage{textcomp}%
\usepackage{xcolor}
\usepackage{hyperref}
\newcounter{notectr}[section]

\renewcommand{\paragraph}[1]{\vspace*{6pt}\noindent\textbf{#1}\;}

\articletype{DATA FOR POLICY CONFERENCE PAPER}
\jname{Data \& Policy}
\jyear{2024}
\jdoi{10.1017/dap.2024.xx}
\begin{document}

\begin{Frontmatter}

\title[Survey]{Case Studies of AI Policy Development in Africa}

% There is no need to include ORCID IDs in your .pdf; this information is captured by the submission portal when a manuscript is submitted. 
% \author[1]{Author Name1}
% \author[2]{Author Name2}
% \author[2]{Author Name3}
\author*[1]{Kadijatou Diallo}
\author*[2]{Jonathan Smith}
\author[3]{Chinasa T. Okolo}
\author[4]{Dorcas Nyamwaya}
\author[4]{Jonas Kgomo}
\author[4]{Richard Ngamita}

\authormark{Diallo \textit{et al}.}

\address*[1]{\orgdiv{Harvard Kennedy School}, \orgname{Harvard University}, \orgaddress{\city{Boston}, \state{Massachusetts}, \country{United States}}. \email{kdiallo10@gmail.com}}
\address*[2]{\orgname{Meta}, \orgaddress{\city{Menlo Park}, \state{California}, \country{United States}}. \email{jajsmith@meta.com}}
\address[3]{\orgdiv{Center for Technology Innovation}, \orgname{The Brookings Institution}, \orgaddress{\city{Washington D. C.}, \country{United States}}. \email{cokolo@brookings.edu}}
\address[4]{\orgname{Equiano Institute}, \orgaddress{\city{Nairobi},  \country{Kenya}}. \email{jonas@equiano.institute}}

\authormark{Diallo et al.}

\received{November 27, 2023}
\revised{January 23, 2024}
\accepted{February 22, 2024}

\keywords{data-ethics, readiness, maturity, governance}
%\keywords[Abbreviations]{EHR, Electronic Health Records; EU, European Union; GDPR, General Data Protection Regulation; HRB, Health Record Bank; ODI, Open Data Institute; PLM, Patients Like Me}

\abstract{Artificial Intelligence (AI) requires new ways of evaluating national technology use and strategy for African nations. We conduct a survey of existing 'readiness' assessments both for general digital adoption and for AI policy in particular. We conclude that existing global readiness assessments do not fully capture African states' progress in AI readiness and lay the groundwork for how assessments can be better used for the African context. We consider the extent to which these indicators map to the African context and what these indicators miss in capturing African states' on-the-ground work in meeting AI capability. Through case studies of four African nations of diverse geographic and economic dimensions, we identify nuances missed by global assessments and offer high-level policy considerations for how states can best improve their AI readiness standards and prepare their societies to capture the benefits of AI.}

\begin{policy}
This research underscores the critical need for reevaluating national technology strategies in African nations amidst the rise of Artificial Intelligence (AI). We identify crucial nuances often overlooked in global assessments by comparing AI readiness indicators and specific barriers African states face. Existing gaps in source data on which assessments are based often ignore significant policy and economic achievements of African states, resulting in indicator scores that belie their progress or, in some cases, global leadership. For example, a number of African states have enacted or introduced legislation to expand public-private collaboration for ethical AI development and introduced unique economic incentives to encourage investment in their native technology sector - actions that can serve as models for other countries. Yet, these actions are often dismissed in existing readiness assessments. Case studies of four diverse African states advocate for a tailored approach to global assessments, emphasizing the importance of addressing local governance challenges and resource availability that vary geographically. The recommendations presented serve as high-level policy considerations, guiding African states toward improved AI readiness. This work contributes to the global conversation on AI policy, fostering a more inclusive and effective approach to the diverse national landscapes in Africa.
\end{policy}

\end{Frontmatter}

\section{Introduction}

The landscape of Artificial Intelligence (AI) technology is undergoing rapid evolution, necessitating a parallel change in the national information technology strategies of African states. This report examines the general AI readiness of four African states, first through the lens of existing assessment frameworks and then through direct investigation of the national readiness of each nation. Our initial analysis shows that current global assessment metrics fail to capture the unique realities of African states in their path to AI readiness. The lack of focus on individual national AI strategies and initiatives within global assessments fails to capture the significant work African states are undertaking to improve their AI readiness.

From a policy perspective, we find that while African countries face a range of difficulties in meeting their technological potential, a few challenges appear wide-spread and critical: weak or insufficient data protection regimes (Mauritius), complications in achieving robust oversight frameworks to support sustained sector growth (Kenya, Egypt), and limited governance and economic resources to devote to technological and workforce development (Angola). Despite the differences in economic development, technical capabilities, and policy prioritization across the continent, we believe countries wishing to accelerate their AI readiness would fare well in improving their outcomes in the areas mentioned above.

While major players such as the U.S., European Union, and China are rapidly developing their AI capabilities, African nations face unique challenges in their path to utilizing AI-enabled technologies to improve their governance and economic outcomes. Limited data access, infrastructure limitations, and scarce political and economic resources challenge African states' ability to develop and utilize AI technologies. Yet, these challenges are not well distinguished in current readiness indicators from the genuine progress on the continent toward AI readiness. This report aims not to provide an exhaustive examination of the overall state of AI readiness across the continent. Instead, it serves as a preliminary assessment of selected countries and suggests a framework for a more in-depth, country-by-country examination. This strategic approach is the first step towards a comprehensive understanding of the nuances and intricacies of preparing African nations for the transformative impact of AI technologies.

\section{Related Work}

A wave of digital readiness policy assessment frameworks precedes the growing literature on AI policy research. These readiness assessments are related to research in organizational theory of change literature, where readiness for change is a "multi-level, multi-faceted construct" involving an informational assessment and organizational commitment and capability to effect change \citep{weiner2009}. Assessing the impact of change readiness in an organizational setting is sparsely researched. Still, at the national level, the gap between practice and theory is more striking, with work primarily focusing on readiness in the cases of conflict resolution \citep{schiff2019} and negotiation \citep{Pruitt2014}. However, because of the proliferation of assessments and their use in driving digital and AI strategy at the national level, it is important to evaluate their effectiveness, and doubly so within the context of the African continent, where there is a "dearth of research that infuses Western conceived theories of management with indigenous ontologies." \citep{Mangaliso2021}

Digital readiness can be assessed at many administrative levels. At the national level, the OECD Digital Government Policy Framework \citep{oecd2020} is one such assessment, focusing on six categories: digital by design, data-driven public sector, government as a platform, open by default, user-driven, and proactiveness. Other frameworks include Gartner's Digital Government Urgency, Readiness and Maturity Assessment \citep{Mickoleit2020}, ITU's e-Government Implementation Toolkit \citep{itu2010}
, the World Bank’s Digital Government Readiness Assessment \citep{worldbank2022}, and Cisco's Digital Readiness Index \citep{cisco2021}. There also exist sub-national policy frameworks, such as the Indian Government's Data Maturity Assessment Framework, used to evaluate the digital readiness of Indian municipalities \citep{india2019}. Some work has been done to evaluate the outcomes of African nations on these assessments, concluding that of the 33 African states present in the 2019 edition of the Cisco Digital Readiness Index, all but one rank below the global average \citep{Assefa2021}. However, many implementations of these frameworks tend to ignore their application in the African continent; for example, the OECD work highlights analysis primarily from member countries, none of which are African \citep{oecd2020}.

Moving beyond digital readiness, there are several AI policy readiness assessments. These assessments, like the previous ones, were found through extensive literature search.
These assessments individually focus on corporate \citep{cisco2023}  \citep{HOLMSTROM2022329}, sub-national \citep{zhang2021ai}, national \citep{maslej2023hai} \citep{rogerson2022oi} \citep{kerry2021brooking} and international policies \citep{kerry2021brooking}. Much of the work on national AI policy analysis has been to define key readiness indicators and track economic and legislative responses to this emerging technology. In these indicators, the work of African nations tends to be overlooked or excluded, even when assessments should indicate progress and, in some cases, global leadership. As an example, in 2021, the Stanford HAI Artificial Intelligence Index Report was published showcasing complete and developing national AI strategy \citep{zhang2021ai}. It only reported two African countries that have announced AI strategies: Kenya and Tunisia, even though both Mauritius and Egypt had already published AI strategies in 2018 and 2019, respectively. It took until the 2023 report \citep{maslej2023hai} for this omission to be corrected.

The 2022 Oxford Insights Government AI Readiness Index \citep{rogerson2022oi} may be the most comprehensive analysis of AI preparedness available today. Once again, it contains gaps in its analysis of the African continent. For example, in the Government AI Readiness Index, Somalia, Equatorial Guinea, and the Sahrawi Republic are all excluded. Additionally, the indicators used omit analyses specific to work being undertaken by African states. The World Bank also conducted an analysis highlighting the challenges of AI in the public sector \citep{worldbank2021}, yet this assessment also included gaps in analysis for Africa, showcasing only a handful of countries on the continent. 

With global AI readiness assessments inadequately capturing African countries, some have advocated for a pan-African AI roadmap focusing on data, infrastructure, education, workforce development, unlocking capital, collaboration, and governance and regulation \citep{diop2023}. One initiative to address African misrepresentation in global assessments was the 2022 creation of the African Observatory on Responsible Artificial Intelligence \citep{africanobservatory2022}. Recognizing "the need to promote African voices, experiences, and value systems in the global debate around responsible AI," it undertook a landscape study of AI policies and use throughout Southern Africa \citep{unesco2022}. This regional analysis was critical and one of the first to focus specifically on Africa, capturing insights for Southern Africa missed in global assessments. This work further highlighted the importance of assessing readiness specific to the continent and as a necessary step to unlocking further policy development for African states.

\section{Steps towards an African AI Policy Evaluation Framework}
Here, we take steps to analyze the applicability of existing national, sub-national, and corporate AI readiness frameworks in the African context. 

We conducted our analysis with the following methodology: we include all AI readiness assessments from the literature review, considering their source, administrative level (one of either National, Sub-national, Municipal, or Corporate), and the number of indicators they include. Each assessment had its indicators manually reviewed and divided into pillars, components, and individual indicators for comparison. Since the current literature did not include an AI-specific municipal-level assessment, the Indian Government City Data Maturity was included as a comparison point for the number of individual indicators. The assessments then were reviewed for their applications of relevant AI policy analysis frameworks at the national level on the level of inclusion of African countries. A country was considered to have coverage if at least one indicator was present in the assessment's application for the given country. Table 1 shows these existing frameworks for AI policy analysis, including their administrative level and number of distinct indicators. 

\begin{table}[htb]
    \caption{AI Policy Analysis Frameworks by Administrative Level}
    \centering
    \begin{tabular}{lp{4cm}lp{2.5cm}}
        \toprule
        \textbf{Framework} & \textbf{Source } & \textbf{Level} & \textbf{Indicators} \\
        \midrule
        Stanford HAI Artificial Intelligence Index & \citep{zhang2021ai} & National & 9 \\
        Brookings AI Policy Analysis & \citep{kerry2021brooking} & National & 11 \\
        Oxford Insights Government AI Readiness \\ Index & \citep{rogerson2022oi} & National & 39 \\
        Stanford HAI Artificial Intelligence Index:\\ U.S. State Analysis & \citep{zhang2021ai} & Sub-national & 5 \\
        Indian Government City Data Maturity \\ Assessment Framework & \citep{india2019} & Municipal & 22 \\
        Cisco AI Readiness Index & \citep{cisco2023} & Corporate & 49 \\
        AI Readiness Framework & (\citep{HOLMSTROM2022329}) & Corporate & 8 \\
        \bottomrule
    \end{tabular}
\end{table}

We consider all 55 states that are members of the African Union \citep{africanunion} and, in Table 2, evaluate how many are included in the applications of relevant AI policy analysis frameworks at the national level. Four of the seven frameworks evaluated contained direct analyses that could be broken down at the national level. One of these was the corporate framework, the Cisco AI Readiness Index, which was sampled from a wide range of nations, while the rest of those with direct analysis were at the national administrative level. Each of the 55 member states of the AU was considered to have coverage only if at least one indicator for the country was present in the published analysis of the AI policy frameworks.

The Oxford Insights' Government AI Readiness Index was found to have the highest coverage, including 52 of the 55 nations studied. We based our analysis and comparison on this assessment since Oxford Insights included the highest number of African states and is generally recognized as a robust global assessment. 

\section{Indicators}
As many of these assessments are aggregated from sources already missing data from Africa, the final assessments further exacerbate the lack of African representation. In this second analysis of the assessments we narrowed our focus to the assessment with the highest existing coverage of African nations, the Oxford Insights analysis. We conducted a manual inspection of the coverage of African countries in individual pillars, components, and indicators (as opposed to in the overall assessment in the previous section). Here we continued to find high rates of missing African country coverage in the underlying data. In some cases, it is unclear if this is the result of omission in the underlying data source due to the inability to find relevant national sources (i.e., a true negative) or due to the omission of the country from the analysis methods (i.e., a potential false negative).

One of the indicators we inspected is in the Technology pillar and Human Capital component: Quality of Engineering and Technology Higher Education. This indicator is based on QS World University Rankings, which only ranks 1500 universities worldwide and only includes African universities in South Africa and Egypt. These kinds of ranking methodologies have been questioned, with the QS ranking, in particular, relying on surveys of academics for 50\% of the 'reputational' dimension \citep{redden2013}. Further investigation is required to determine if there may be any bias specifically against African countries in these rankings. However, other similar rankings, such as the Academic Ranking of World Universities (ARWU), which are primarily metric-based rather than survey-based, include greater representation of universities within the African continent. For example, in 2023, ARWU included five African states in its ranking of the top 1000 institutions \citep{shanghai}. Other metric-based assessments, such as the AI Research Papers Indicator- Innovation Capacity, also tend to show higher and more diverse African representation.

A second indicator that was inspected is the Open Data indicator: Infrastructure and Data and Data Availability component. The Open Data indicator contributes to defining AI readiness by assessing how much data is available to train AI models. It is constructed based on the Global Data Barometer, which covers 109 countries. It appears to be missing data for a large number of African countries. The underlying data source from the Global Data Barometer website contains data on 23 African states or just 41\% of the continent. However, all countries without known data receive a 0 as part of the assessment rather than a 'Not Available' mark, which contributes to lowering the readiness score of a majority of nations and the continent as a whole.  

It is apparent that even for frameworks with high coverage of African states in direct analysis, underlying data gaps remain that can result in disproportionately lower readiness assessment scores for the continent. 

\begin{table}[htb]
    \caption{Inclusion of African Nations in AI Policy Analysis}
    \centering
    \begin{tabular}{lccc}
        \toprule
        \textbf{Framework} & \textbf{Direct } & \textbf{ Level} & \textbf{African States / \% Coverage} \\
        \midrule
        Stanford HAI AI Index (2023) & Yes & National & 16 / 29\% \\
        Brookings AI Policy Analysis & Yes & National & - / - \\
        Oxford Insights AI Readiness Index & Yes & National & 52 / 94\% \\
        Stanford HAI AI Index: U.S. State Analysis & Yes & Sub-national & - / - \\
        Indian Govt. City Data Maturity Framework & No & Municipal & - / - \\
        Cisco AI Readiness Index & Yes & Corporate & 1 / 1.8\% \\
        AI Readiness Framework & No & Corporate & - / - \\
        \bottomrule
    \end{tabular}
\end{table}

 After analyzing existing AI Policy readiness assessments, it becomes clear that in the context of Africa, more targeted qualitative assessments should be undertaken to fully understand the gaps and directions for policymakers to focus on. In the following case studies, we will review three qualitative indicators we believe would better capture the African state's true AI readiness and progress. Assessments can more accurately reflect a state's true reality by reviewing available laws and strategy documents and considering the context of both digital and AI readiness alongside economic and external factors. 

\section{Case Studies: African States and AI Policies}
In selecting our case studies, we sought states of different maturity concerning national technology development, governance, and economic diversity. While many reports focus on the continent's larger economies, like Nigeria and South Africa, or states with robust technological and innovative capacity, like Rwanda, we sought to highlight less-researched states that have made efforts in recent years to improve their digital infrastructure and capabilities to better exploit AI for development growth. In our case assessments, we look at the full body of legislative activity and documents concerning broader issues of technology and innovation. The countries we examine have dedicated legislative frameworks for digital innovation, such as a national strategy for AI (Egypt) or a multi-year strategic plan for digitization of services (Mauritius, Kenya, Angola). We analyze both flagship legislation - such as strategic plans - and the broader ecosystem of activity each country is engaged in to meet its technological development goals, such as laws around data protection and improving investment in their technology sectors.

These states present a unique opportunity to showcase how focusing on a specific policy improvement can drastically improve their overall AI readiness scores on assessments and in practice. Using Oxford Insights \citep{rogerson2022oi} as a baseline, with the exception of Angola, these states occupy middling global but high Africa-region rankings,  suggesting that while they have room for improvement, those improvements are relatively achievable with a focus on select critical areas. For states facing a myriad of governance and economic challenges across domains, the ability to target a specific policy with high returns and a chance of success is crucial. For these reasons, our analysis pinpoints a narrow policy focus for each case study, arguing that improvement in that area offers greater return for the country.  

A particularly interesting finding we surfaced further highlights the limitations of global indicators for Africa. In the case of Mauritius, while the country was the highest-ranked African state and scored decently on the Data and Infrastructure Pillar, our examination of Mauritius' data policies revealed weaknesses in its regulation that we argue are not accurately captured in its Data Pillar score. Since data is a crucial component of any AI readiness, a lack of rigorous examination into the efficacy and robustness of a country's data policies is short-sighted and misleading. Conversely, Kenya scored lower on its governance score, but an analysis of its national strategy and initiatives highlights a government that's focused its considerable political and economic currency on creating an enabling environment for AI usage. Here, we see once more that the lack of contextual examination within these indicators does not fully capture the status of African states. 
\begin{table}[htb]
    \caption{Case Study Country Oxford Insights Government AI Readiness Assessment}
    \centering
    \begin{tabular}{lcccccc}
        \toprule
        \textbf{Country} & \textbf{ Score} & \textbf{Global Rank} & \textbf{African Rank} & \textbf{Government} & \textbf{Technology} & \textbf{Infrastructure}\\
        \midrule
        Mauritius & 53.38 & 57 & 1 & 68.65 & 31.16 & 60.34 \\
        Egypt & 49.42 & 65 & 2 & 63.46 & 36.07 & 48.72 \\
        Kenya & 40.36 & 90 & 6 & 40.36 & 28.76 & 51.95 \\
        Angola & 24.77 & 163 & 38 & 21.33 & 14.98 & 37.99 \\
        \bottomrule
    \end{tabular}
\end{table}

\subsection{Mauritius: Strengthening Data Access and Protection}

While a small island nation, Mauritius has made notable progress in developing comprehensive policies to govern its technology usage. The Mauritian government recognizes the potential of AI in improving economic and social outcomes and transforming the state into a "smart" island equipped for the future. The 2018 release of its Digital Mauritius 2030 Strategic Plan laid the foundation for addressing the country's AI readiness \citep{mauritius2018}. It ushered in the convening of the Mauritius Artificial Intelligence Council and addressed workforce skill development to meet the nation's talent needs. It tied AI deployment to improving development outcomes and making governance more effective and responsive to citizen needs. Four years after the release of the digital strategic plan, in 2022, Mauritius was the highest-ranked African country included in the Oxford Government AI Readiness Index, scoring 53.38 and placing 57th out of 181 countries \citep{rogerson2022oi}. 

The 2017 Data Protection Act of Mauritius, enacted in January 2018, serves as a comprehensive framework governing the access and use of personal data by individuals and public bodies. The Act is locally focused and applicable to entities based in Mauritius or using Mauritian-based equipment for data processing. Oversight is provided by the Data Protection Office, led by an independent Data Protection Commissioner, tasked with implementing the Act, issuing guidelines, and initiating investigations. However, the Act lacks clarity on prosecutorial powers, raising questions about its enforceability, failing to address regulation for private firms, and risks becoming obsolete without a clearly defined revision of powers.   

A significant limitation is the Act's exclusive emphasis on individuals and public entities, neglecting regulations for private firms. This gap raises concerns about the adequacy of protection for personal data handled by private organizations, regardless of location, especially considering the increasing role of the private sector in data processing as it falls short in addressing transboundary data processing. The need for authorization from the DPC and proof of data protection safeguards for data transfer outside Mauritius raises questions about the feasibility and efficiency of such processes. As data is foundational to AI readiness, how a country mandates the compilation, storage, and protection of public data is an important indicator of how it can responsibly meet the future technological landscape. In this case, Mauritius's Data Pillar score reveals its need for improvement.  

The Act lacks provisions for data subjects to exercise their rights effectively, and there is a notable absence of regulations regarding "data portability," a crucial element addressed in other data protection regulations such as the EU's Global Data Protection Regulation (GDPR). Overall, there is a concern about the rigor of implementation of the Act, especially within the government itself. As an example, civil society notes the government's access to CCTV data and the recent mandatory re-registration of mobile data users being completed without clarity for how the state stores this data or for what purposes it is used \citep{UN2023}.   

Additionally, we found the lack of strategy for updating the state's data protection regulations to be a barrier to effectively making use of its capabilities. There is no clearly defined process to ensure the robust evolution of data protections to meet the demands of a rapidly changing landscape, particularly as the country bolsters its capacity to provide more e-services. For example, in its current iteration, the Act does not provide clear guidance on the authority of the DPC concerning cross-boundary or multi-jurisdictional virtual financial transactions within Mauritius. As of the submission of this paper, we did not find any indication of Mauritius amending the Act to account for these concerns. While technology evolves faster than legislation for any country, barely five years since enactment, the Act already appears to be falling short of meeting the data protection needs of a rapidly changing sector.

\subsection{Egypt: Improving Investment in AI}

Egypt was fairly early in creating a national taskforce to examine the implications of this new technology. A year after the EU's GDPR established its own process, the Egyptian National Council for AI (NCAI) was formed in 2019. Egypt's National AI Strategy is framed around four pillars: government, development, capacity building, and international relations. Notably, the strategy allocates a budget of USD 318M for digital transformation projects, underlining the government's commitment to advancing AI initiatives. The strategy also emphasizes responsible and ethical AI practices, with plans to incorporate a dedicated track within the NCAI for AI ethics. The Egyptian Charter for Responsible AI, implemented in April 2023, complements these efforts by providing guidelines to enhance stakeholder awareness regarding the risks and limitations of AI. These advancements have also been appreciated by neighbouring states, leading to Egypt's election as chair of the Arab AI working group \citep{egyptaichair}.

In AI readiness terms, Egypt was considered the second highest-ranked African country in the Oxford Government AI Readiness Index, scoring 49.42 and placing 65th out of 181 countries \citep{rogerson2022oi}. The country's lowest performance was seen in its  Technology pillar, despite Egypt's ongoing investment in improving its ICT infrastructure and attracting investments to develop its technology sector and capabilities. Egypt produces the most AI research papers of any African country and records the highest R\&D spending on the continent.

In 2022, the MCIT implemented legislative changes to boost the IT and electronics industries and foster the growth of ICT startups. Measures included facilitating the establishment of companies through digital notices, allowing virtual companies without physical premises, and reducing the minimum capital for one-person companies. Notably, the mobile phone industry received incentives to attract investments, and a cooperation protocol was signed to create an enabling environment for technology startups. The 2022 Egyptian Economic Conference yielded recommendations focusing on stimulating research and development, enhancing language skills programs, establishing an economic zone for the IT industry, and extending financial incentives for SMEs in the technology sector. These developments underscore Egypt's proactive approach to fostering innovation and technological growth while adapting to the evolving global landscape - initiatives that are not fully reflected in its assessment. Here, once more, we see the limits of current global assessments in not considering proactive policies African states are taking to improve their AI readiness in the context of their resource constraints. 

\subsection{Kenya: Creating a Robust Enabling Environment}

The Kenyan government has proactively embraced AI as a catalyst for national development. In 2018, a dedicated AI taskforce was established with the explicit goal of formulating a comprehensive national AI strategy. This initiative sought to position Kenya as a global AI hub and foster the widespread adoption of AI technologies across key sectors. Notably, this taskforce laid the groundwork for the subsequent launch of the 10-year \href{https://cms.icta.go.ke/sites/default/files/2022-04/Kenya%20Digital%20Masterplan%202022-2032%20Online%20Version.pdf}{National Digital Master Plan 2022-2032} by the Ministry of Information, Communication, and Technology (ICT).

In AI readiness terms, Kenya is the sixth highest-ranked African country in the Oxford Insights Government AI Readiness Index, scoring 40.36 and placing 90th out of 181 countries \citep{rogerson2022oi}. Kenya's score is bolstered by its performance in the Data and Infrastructure pillar. Yet, its low Government Pillar belies the government's continued efforts to create an enabling environment for the effective and responsible development of AI integration.

The National Digital Masterplan, spanning from 2022 to 2032, marks a significant milestone in Kenya's technological trajectory. Aligned with the country's moniker "Silicon Savannah," the master plan ambitiously targets the transformation of priority sectors such as agriculture and healthcare through the strategic integration of AI. This underscores Kenya's commitment to leveraging advanced technologies to address critical challenges and drive sustainable development.

In a move signaling the government's continued commitment to fostering a conducive environment for AI, President William Ruto issued a directive in September 2023 tasking the ICT Ministry with the development of comprehensive Artificial Intelligence legislation. The objective of the directive was to create an enabling regulatory framework that enhances digital competitiveness and facilitates responsible AI deployment.

The 2022 AI strategy report presented by the taskforce laid out key recommendations for effective AI governance and regulation. These include the establishment of ethical guidelines for AI development and usage, the implementation of robust measures to safeguard privacy and security, the promotion of AI skills and research capacity, the facilitation of partnerships among government, academia, and industry, and the development of policies to mitigate risks and biases inherent in AI systems.

\subsection{Angola: Laying a Solid Foundation}

While we found assessment indicators tended to underestimate countries in areas of improvement, indicators were more in line with identifying states with poor readiness scores across the board. Angola, one of the world's largest oil producers, finds itself among Africa's least AI-ready economies, with Oxford Insights finding it the thirty-eighth-ranked African country, scoring 24.77 and placing 163rd out of 181 countries \citep{rogerson2022oi}. 

In 2021, the Minister of Telecommunications, Information Technologies, and Media, during the Forum on Digital Transformation, emphasized the potential of AI in driving the country's digital evolution. While plans for digital transformation are underway, exemplified by the signing of memoranda of understanding with the UAE for the Digital Angola 2024 strategy \citep{digitalangola}, Angola's currently has a limited focus on AI readiness. In this way, Angola, among all the case study countries, shows the effectiveness of the current AI readiness assessments.

Although Angola has established data privacy and protection regulations, including the Data Protection Law, Electronic Communications and Information Society Services Law, and Protection of Information Systems and Networks Law, challenges persist \citep{angolalaws}. The nation's heavy reliance on oil, a predominant economic sector, constrains the government's ability to prioritize AI development. Notable initiatives within the oil sector, such as the establishment of a new R\&D center by Sonangol, the national oil company, underscore the concentration of resources in this sector. Given the lack of AI innovation and regulation in Angola, the country's focus on building human capital in gas, mining, and renewable energy diverts financial and legislative attention away from making progress toward strengthening its digital economy.

Governance challenges, particularly those associated with the corruption experienced within the Dos Santos regime, have further strained government resources. Additionally, pressing socioeconomic needs, such as 44\% of citizens lacking access to clean water \citep{casimiro2019} and 48.2\% having limited access to electricity \citep{worldbank2024}, are understandably of more significant concern than AI development. Angola struggles with funding public education, which has led to instability within its higher education institutions. Within the past two years, faculty have gone on strike three times \citep{lazaro2023}. This lack of funding highlights a significant gap in Angola's technological landscape, leaving advanced research within its early stages and leading to a lack of sufficient scholars, institutes, and startups dedicated to AI advancement.

To make progress towards becoming an AI-ready economy, our recommendations for Angola center on harnessing local AI talent by recruiting experts for taskforces to guide governmental efforts on AI regulation. Simultaneously, investments in electrical and data infrastructure are vital to enhance research capabilities. Emulating the model of the new national oil company R\&D center, there is a need for strategic investments in building AI institutes to facilitate the scaling of local AI research. Furthermore, integrating STEM training into primary and secondary school curricula and investing in post-secondary training and upskilling initiatives are crucial steps to nurture the human capital necessary to sustain Angola's emerging AI industry. By adopting these recommendations, Angola can pave the way for a more robust and inclusive AI ecosystem. Overall, this case study provides potential pathways for countries with socioeconomic similarities to Angola to approach AI development and governance.

\section{Future Policy Indicators}

The last few years have seen an influx of global assessments to measure countries' AI readiness. However, when evaluating the AI readiness of African states, it becomes evident that current global assessments often inadequately depict the nuanced landscape of these nations. The reliance on data sources that may exclude or misrepresent African realities, coupled with indicator definitions that fail to encompass the full spectrum of governmental and economic activities, results in rankings that do not adequately reflect the triumphs and challenges faced by African states.

To address these challenges, it is crucial for global assessments to reassess their indicator measurement strategies and data sources. Assessments should tailor their approach to Africa's distinct economic requirements, taking into account capacity, resources, and the continent's history of technological advancements. Also, considering the historical lack or poor quality of empirical data specific to African development, a more comprehensive and innovative approach that involves integrating diverse sources, resonating with the unique context and aspirations of African nations, is needed. For instance, a closer examination of national policies can provide a more accurate picture of the legislative (Mauritius), economic (Egypt), and technological (Kenya) environments and aspirations for African states. To that end we propose several indicators that should be taken into account in future assessments, along with sources that can substantiate these indicators.

\begin{enumerate}

    \item \textbf{Foundational Technology Infrastructure.} We define foundational technology infrastructure as national policies or strategies, either existing or in development, that focus on creating baseline technological engagement and deployment: data protection legislation, information and communication technology development, digital governance, and a technically skilled workforce. Existing sources for this, as highlighted in the Indicators section, are lacking, despite the inclusion of related indicators in current assessments. To better complete these indicators we suggest working with individual nations to unearth currently adopted metrics for such indicators. For example, in the Angola case study we see the prioritization of STEM training inclusion in school curricula that could be used as an indicator for improving the technical skill of the workforce.
    \item \textbf{Speed of Technological Innovation Adoption}. This indicator recognises Africa's speed in adopting new technology by measuring the pace of adoption, rather than simply the current state. It recognizes the continent's innovation prowess and leadership in adopting new technologies, such as cellular phones and mobile banking. By considering the second order effect of rate of change in existing innovation, data, and infrastructure indicators there can be additional context provided to assessors.
    \item \textbf{Enabling Environment for AI Legislation.} This indicator uniquely highlights the legislative landscape's nuances and interest in AI-specific regulation by scrutinizing government publications on technology and digitization. Its distinct contribution lies in the focused examination of specific AI-specific policy areas, such as existing or proposed legislation, regulation, ethical development, workforce training, among others. While the Stanford HAI Index \citep{maslej2023hai} includes four indicators related to this kind of policy making it has very low coverage of the African context. New sources like the Open African Laws \citep{openafricalaws} programme can increase coverage with their focus on the African context. Additionally, an indicator related to the identification of specific agencies or working groups dedicated to AI, underscoring a proactive commitment to enhancing AI readiness would be another way to bridge this gap. This approach is likely to address Africa's unique challenges in policy development and regulatory frameworks by showcasing a comprehensive understanding of the specific areas requiring attention and development in the AI domain.
    \item \textbf{Economic and Development Context.} This indicator uniquely addresses the necessity of considering the diverse development timelines, economic capacities, and developmental stages of African nations during AI readiness assessments. By strategically acknowledging challenges such as lower Gross Domestic Product (GDP), and Human Development Index (HDI), it ensures a comprehensive understanding of Africa's readiness for AI, which can lead to targeted solutions for specific economic contexts. The indicator goes beyond the legislative examination and narrow economic indicators of existing assessments. For example, the Stanford HAI Index \citep{maslej2023hai} includes indicators three economic indicators: federal budget for non-defense AI R\&D, defense budget requests for AI-specific research, and government AI-related contract spending. In the Oxford Insights \citep{rogerson2022oi} assessment there are some broader indicators such as company R\&D spending, VC availability, and value of trade in ICT goods and services, however this can miss context such as different funding models (non-VC) or non-trade economic power that can be highlighted with the indicator as proposed here.
    \item \textbf{Geopolitical Context.} This indicator offers a lens through which to understand the historical and geopolitical factors influencing AI readiness in Africa. Existing indicators consider government capacity and effectiveness, however they do not consider the broader context driving those indicators. As shown in the case study of Egypt, where the country is leading a multi-national working group on AI, surrounding countries can benefit from strong regional players. Considering the preparedness and government effectiveness of neighbouring states or trade partners can be an additional indicator of AI readiness.
   
\end{enumerate}

\section{Considerations for States on Improving AI Readiness}

For African states, while current indicators may not accurately communicate their progress on AI readiness, they confirm that the continent faces challenges in improving its AI capabilities. While the difficulties confronting African states vary in their manifestations, our analysis of the selected case studies suggests a few significant shared challenges across the continent, the improvement of which would significantly accelerate state and continental AI readiness. The following considerations address high-level policy improvements that states can pursue that align with existing country and continent-wide initiatives, notably the African Continental Free Trade Area (AfCFTA) and initiatives currently underway by the African Union. These targeted improvements have the potential to positively impact diverse aspects of each country's socioeconomic landscape and better reflect the continent's AI readiness to the world. 

\begin{enumerate}
    \item \textbf{Improve Data Infrastructure and Protections.} The building block of successful AI-enabled technologies is access to quality data. If states are to support nascent AI developments, they must improve the quality of their public data and provide clear guidelines for data access meant for private and public usage. This needn't fall entirely on the state; collaboration with civil society groups, universities, or public research centers can allow for more specialized, robust, and responsible collection and maintenance of public data.
    \item \textbf{Regional and Continental Cooperation.} Understanding the full potential of AI and identifying the responsible development and usage of these technologies is daunting for any country. African states need not go it alone. An advantage that African states have are the regional and continental-wide communities they already collaborate with in other critical policy areas. Sharing best practices on a bilateral or regional basis can facilitate greater progress for the continent and position it as a global leader in defined areas of the field. A more coordinated approach by African states can also improve their global competitiveness to influence technological policy and shape standards and indicators that better reflect their realities.  
    \item \textbf{Public Engagement and Development of Human Capital.} Regardless of the policies African states pursue, their success depends on their ability to utilize their human capital resources effectively. As states seek to draft or improve their national data and AI strategies, they should implement clear opportunities for public input from all sectors of society: from industry experts, civil society, and the general public. Implementing outlets for public comments or requests for opinions can result in more robust, technically sound, and socially responsible frameworks and legislation. Equally important, states should incorporate workforce development and technical skill-building provisions within their respective AI and data governance strategies.
\end{enumerate}

\section{Conclusion}

In this work we analyze the existing AI readiness of Mauritius, Egypt, Kenya, and Angola, focusing on existing assessments frameworks and their relevance to African countries. We find that current assessments fail to adequately capture the readiness of African states. Through case studies of each country, we identify limitations and shortcomings of current AI readiness assessments vis-à-vis Africa and put forth recommendations for revising assessment methodologies to better account for the continent's economic and policy contexts, incorporating more qualitative analysis of countries' technology strategies and governance initiatives. This provides a necessary building block for elevating AI readiness around the world.

\begin{Backmatter}

\paragraph{Provenance}
This article was submitted for consideration for the 2024 Data for Policy Conference to be published in Data and Policy on the strength of the Conference review process.

\paragraph{Acknowledgments}
TBD

\paragraph{Funding Statement}
None

\paragraph{Competing Interests}
None

\paragraph{Data Availability Statement}
Data on the AI readiness analysis by African nations will be available online on GitHub at the conclusion of reviewing.

\paragraph{Author Contributions}
K.D. and J.S. designed the study, retrieved the data, conducted analysis, wrote the first draft, and approved the final version of the manuscript. C. T. O., D  N., J. K., and R. N. conducted the case study analyses, wrote the first case studies, revised the manuscript, and approved the final version of the manuscript.

\paragraph{Additional Material}
None

\bibliographystyle{apalike}
\bibliography{paper.bib}

\end{Backmatter}

\end{document}